\documentclass[10pt]{amsart}
\usepackage{times}

\usepackage[all]{xy}

\SelectTips{cm}{}
\usepackage{amsbsy,mathrsfs,latexsym,amssymb,amsmath}

\usepackage{amsmath}
\usepackage{bbm}

\def\Set{{\mathsf{Set}}}
\def\Nom{{\mathsf{Nom}}}
\newcommand{\Srt}{\mathit{Srt}}
\newcommand{\Op}{\mathit{Op}}
\def\kat#1{{\mathscr{#1}}}

\def\K{\kat{K}}

\def\A{\kat{A}}

\def\X{\kat{X}}
\def\Alg#1{{\mathsf{Alg}}(#1)}
\def\eps{\varepsilon}
\def\fp{{\mathit{fp}}}
\def\Lan#1#2{{\mathrm{Lan}}_{#1} #2}
\newcommand{\supp}{\mathsf{supp}}
\def\fbEnd#1{{\mathsf{End}}_{{\mathit{fb}}}(#1)}
\def\fbMnd#1{{\mathsf{Mnd}}_{{\mathit{fb}}}(#1)}

\newcommand{\app}{\mathsf{app}}
\newcommand{\Fr}{\mathsf{Fr}}
\newcommand{\rules}[2]{\mbox{$\frac%
    {\mbox{\normalsize \rule[-5pt]{0pt}{14pt} $#1$}}
    {\mbox{\normalsize \rule[0pt]{0pt}{10pt}$#2$}}$}}
\newcommand{\sfresh}{\ensuremath{\mathrel{\mathpalette\sfreshs\approx}}}
\newcommand{\sfreshs}[2]{\ooalign{$\hfil#1\mkern1mu/\mkern-6mu/\hfil$\crcr$#1#2$}}

\newcommand{\acal}{\mathcal{ A}}

\newcommand{\ccal}{\mathcal{ C}}

\newcommand{\ical}{\mathcal{ I}}

\newcommand{\ncal}{\mathcal{ N}}

\newcommand{\ucal}{\mathcal{ U}}

\newcommand{\fbb}{{\ensuremath{\mathbb{ F}}} }
\newcommand{\ibb}{{\ensuremath{\mathbb{ I}}} }

\newcommand{\lbb}{{\ensuremath{\mathbb{ L}}} }

\newcommand{\mbb}{{\ensuremath{\mathbb{ M}}} }
\newcommand{\nbb}{{\ensuremath{\mathbb{ N}}} }

\newcommand{\tbb}{{\ensuremath{\mathbb{ T}}} }

\newtheorem{theorem}{Theorem}[section]
\newtheorem{definition}[theorem]{Definition}

\newtheorem{proposition}[theorem]{Proposition}
\newtheorem{remark}[theorem]{Remark}

\newtheorem{example}[theorem]{Example}

\begin{document}

\title{Algebraic Theories over Nominal Sets}

\author{Alexander Kurz}
\address{Department of Computer Science, University of Leicester, United Kingdom}
\email{kurz@mcs.le.ac.uk}
\author{Daniela Petri{\c s}an}
\address{Department of Computer Science, University of Leicester, United Kingdom}
\email{petrisan@mcs.le.ac.uk}
\author{Ji\v{r}\'{i} Velebil}
\address{Faculty of Electrical Engineering,
   Czech Technical University in Prague}
\email{velebil@math.feld.cvut.cz}
\thanks{The third author acknowledges the support of the grant MSM6840770014
of the Ministry of Education of the Czech Republic.}
\date{}
\maketitle
\thispagestyle{empty}

\begin{abstract}
\noindent We investigate the foundations of a theory of algebraic
data types with variable binding inside classical universal
algebra. In the first part, a category-theoretic study of monads
over the nominal sets of Gabbay and Pitts leads us to introduce
new notions of finitary based monads and uniform monads. In a
second part we spell out these notions in the language of
universal algebra, show how to recover the logics of
Gabbay-Mathijssen and Clouston-Pitts, and apply classical results
from universal algebra.
\end{abstract}

\section{Introduction}

\noindent The nominal sets of Gabbay and
Pitts~\cite{gabb-pitt:lics99} give an elegant and powerful
treatment of variable binding which is, on the one hand, close to
informal practice and, on the other hand, lends itself to rigorous
formalisation in theorem provers or programming languages. Nominal
sets have been extraordinarily successful as witnessed by
a wide range of work.

\medskip\noindent Closely related, albeit less developed, are the models of
variable binding based on presheaf categories $[\ical,\Set]$.
These are categories of functors $\ical\to\Set$ where the indexing
category $\ical$ consists of contexts (=sets of free variables)
and maps between them (such as weakenings and renamings). This
started with \cite{fpt:lics99,hofmann:lics99} and was axiomatised
in \cite{tana-powe:variable-binding} to treat different $\ical$ in
a uniform way. We focus on the indexing category $\ibb$ associated
with nominal sets (more below) and leave the general theory for
future work.

\medskip\noindent  This paper presents the foundations of a theory of
algebraic data types with variable binding. We do this inside
standard many-sorted universal algebra. In particular, the logics
arising are (fragments of) the standard ones based on equational
logic. This enables us to leverage the existing theory of
universal algebra and we illustrate this by transferring two
classical theorems to nominal sets: Birkhoff's variety theorem (or
HSP-theorem) characterising equationally definable classes of
algebras; and the quasivariety theorem characterising
implicationally definable classes (Section~5).

\medskip\noindent We proceed in the following way. Although the category
$\Nom$ of nominal sets is not equationally definable itself, it
embeds in a canonical way  into a presheaf category $[\ibb,\Set]$,
sorted over contexts. Like any presheaf category, $[\ibb,\Set]$ is
a many-sorted variety, ie equationally definable. Thus, over
$[\ibb,\Set]$, universal algebra can be done in the usual way, by
adding operations and equations. Transferring this back to nominal
sets, it turns out that the logic thus obtained is more general
than what is usually intended when working with nominal sets. The
reason is that over $[\ibb,\Set]$ we have access to individual
contexts and can define theories which do not treat contexts in a
uniform way. This is repaired by introducing uniform theories. We
then show that the (quasi)variety theorems specialise to uniform
theories (Section 4).

\smallskip\noindent Three points are worth noting:

\par\smallskip\noindent\textbf{Nominal sets and sets-in-context.} There has been some debate on whether nominal sets or sets-in-context are preferable. We illustrate how both have their advantages. On the one hand, our concept of a uniform theory originates from Gabbay's discovery \cite{gabbay:hsp} that classes of algebras over nominal sets (in the sense of \cite{gabbay:hsp}) are closed under abstraction (Definition~\ref{def:abstraction}). On the other hand, the sets-in-context approach of $[\ibb,\Set]$ allows us to use universal algebra directly and we obtain Gabbay's HSP-theorem and novel variations as a corollary of the classical theorems.

\par\smallskip\noindent\textbf{Category theory (CT).} Category theory appears in this work for several reasons. First, CT offers a widely accepted notion of algebraic theory over a category, namely that of a monad. Thus, an account of algebraic theories over nominal sets ignoring monads would be incomplete. Second, the relationship between nominal sets and sets-in-context is best formulated in CT, see for example the crucial `transport theorems' of Section~\ref{sec:transport}. Third, CT allows for proofs at the right level of abstraction, thus providing more general results and opening new directions, some of which we will discuss in the conclusions. 

\par\smallskip\noindent\textbf{Fb-monads.} The categorical analysis of monads on nominal sets leads us to add fb-monads to the powerful toolbox of CT in computer science. They arise because monads on nominal sets are too general to remain in the realm of equational logic and universal algebra.
Whereas fb-monads are precisely those monads which can be
presented in universal algebra. Moreover, they can be transported
from nominal sets to $[\ibb,\Set]$ and back: Loosely speaking,
universal algebra does not see the difference between the two
categories.

\smallskip\noindent The structure of the paper is as follows. Section 3 studies monads on nominal sets and $[\ibb,\Set]$ and introduces fb-monads and uniform monads. Section 4 develops universal algebra
over $[\ibb,\Set]$ and gives a syntactic description of the notion
of uniform theory. Section 5 applies these results to algebras
over nominal sets and shows that the work of Gabbay and Mathijssen
\cite{gabb-math:nomuae} and Clouston and Pitts
\cite{clou-pitt:nom-equ-log} fit in our framework.

\section{Preliminaries}\label{sec:preliminaries}

\noindent\textbf{Notations.} If $\A$ is a small category and $\K$
an arbitrary category the functor category $[\A,\K]$ has as
objects functors from $\A$ to $\K$ and as morphisms natural
transformations between functors.

\medskip\noindent For an endofunctor $L$ on a category $\A$, we consider the
category of \emph{$L$-algebras}, denoted by $\Alg{L}$, whose
objects are defined as pairs $(A,\alpha )$ such that $\alpha: LA
\rightarrow A$ is a morphism in $\A$. A morphism of $L$-algebras
$f: (A, \alpha) \rightarrow (A',\alpha ')$ is a morphism  $f:A
\rightarrow A'$ of $\A$ such that $f \circ \alpha=\alpha ' \circ
Lf$.

\medskip\noindent If $\A$ is a category and $\mbb=(M,\mu,\eta)$ is a monad on
$\A$ then $\A^\mbb$ denotes the category of Eilenberg-Moore
algebras for the monad $\mbb$. These are algebras for $M$ that
behave well with respect to the multiplication and unit of the
monad, see~\cite{cwm} for a precise definition.

\medskip\noindent  If $L$ is either a functor or a monad we use the ad-hoc
notation $L$-$\mathit{Alg}$ for algebras for $L$.

\medskip\noindent If $S$ is a set and $A$ an object in a cocomplete category
$\K$, $S\bullet A$ denotes the copower, that is, the coproduct of
$S$-copies of $A$.

\medskip\noindent\textbf{Universal algebra (UA) and
UA-presentations.} A signature $(\Srt,\Op)$ in the sense of UA, or
a \emph{UA-signature}, is given by a set $\Srt$ (of sorts) and a
set $\Op$ of operation symbols $\mathit{op}:w\to s$ where $w$ is a
finite word over $\Srt$ and $s\in\Srt$. A \emph{UA-theory}
$\langle \Srt, \Op, E\rangle$ is given by a UA-signature and a set
$E$ of equations and $\Alg{\Srt,
  \Op, E}$ is the class of its models. If a category $\A$ is
isomorphic to $\Alg{\Srt,\Op, E}$ we say that $\langle \Srt, \Op,E
\rangle$ is a \emph{UA-presentation} of $\A$ and call $\A$ a
\emph{variety}. A variety $\A$ comes with a forgetful functor
$U_\A:\A\to\Set^\Srt$, which has a left-adjoint $F_\A$. 
%

\medskip\noindent\textbf{Monads.}
Any adjunction $F\dashv U:\K\to\X$ gives rise to a monad
$\tbb=UF$, which in turn determines the category $\X^\tbb$ of
algebras for the monad. If $\tbb$ is finitary (=preserves filtered
colimits \cite{ar}) and $\X=\Set^\Srt$, then $\X^\tbb$ is a
variety. Conversely, any variety $\A\cong\Alg{\Srt,\Op, E}$ is
isomorphic to $(\Set^\Srt)^\tbb$ where $\tbb=U_\A F_\A$ is a
finitary monad. We  say that $\langle \Srt,\Op, E\rangle$ is a
UA-presentation of the monad $\tbb$.

\medskip\noindent\textbf{Nominal Sets.} We consider a countable
set $\ncal$ of names and the group $\frak{S}(\ncal)$ of finitely
supported permutations on $\ncal$ (that is permutations that fix
all but a finite set of names). Let $\cdot:\frak{S}(\ncal)\times
X\to X$ be a left action of the group $\frak{S}(\ncal)$ on a set
$X$. We say that a finite subset $S\subset\ncal$ supports an
element $x$ of $X$, if for any permutation $\pi\in\frak{S}(\ncal)$
that fixes the elements of $S$ we have $\pi\cdot x=x$. A
\emph{nominal set} is a left action $(X,\cdot)$ such that any
element of $X$ is supported by a finite set.

\medskip\noindent For each element $x$ of a nominal set there exists a
smallest set, in the sense of inclusion, which supports $x$. This
set, denoted by $\supp(x)$, is called the \emph{support} of $x$.
We say that $a\in\ncal$ is \emph{fresh} for $x$ if
$a\not\in\supp(x)$.

\medskip\noindent A morphism of nominal sets $f:(X,\cdot)\to(Y,\circ)$ is an
\emph{equivariant} map between the carrier sets: $f(\pi\cdot
x)=\pi\circ f(x)$ for all $x\in X$. Let $\Nom$ be the category of
nominal sets and equivariant maps.

\medskip\noindent\textbf{Nominal sets and  the functor category
$[\ibb,\Set]$.}  The notion of support equips $\Nom$ with a
forgetful functor $U$, which in turn generates the variety
$[\ibb,\Set]$ and the embedding $\Nom\to[\ibb,\Set]$. Here, $\ibb$
is the category whose objects are finite subsets of $\ncal$ and
morphisms are injective maps. The underlying discrete subcategory
is denoted by $|\ibb|$.

\medskip\noindent To define $U:\Nom\to[|\ibb|,\Set]$, we let, for a nominal
set $X$, $UX(S)$ be the set of elements of $X$ supported by $S$.
$U$ has a left adjoint $F:[|\ibb|,\Set]\to\Nom$.\footnote{The
nominal sets $FY$ are the strong nominal sets of
\cite{tzevelekos:lics07}.} Let $\tbb$ denote the monad on
$[|\ibb|,\Set]$ generated by $F\dashv U$. The category of
Eilenberg-Moore algebras for the monad $\tbb$ is equivalent to
$[\ibb,\Set]$. The adjunction $F\dashv U$ is not monadic, but
rather of \emph{descent type}: this means that the comparison
functor $I:\Nom\to[\ibb,\Set]$ is full and faithful.

\begin{equation}\label{equ:nom-triangle}
  \xymatrix{
    \Nom \ar@/_.7pc/[rr]_{I}\ar@/^.7pc/[ddr]^{U}&\bot& {[\ibb,\Set]}\ar@/_.7pc/[ll]_{I^*}
    \ar@/^.7pc/[ddl]^{U^\tbb}\\
    & & \\
    &[|\ibb|,\Set]\ar@/^.7pc/[uur]^{F^\tbb}\ar@/^.7pc/[uul]^{F}\POS!R(.7),\ar@(ur,dr)^{\tbb} &
    }
\end{equation}
$\Nom$ is equivalent to the full reflective subcategory of
$[\ibb,\Set]$ consisting of pullback preserving functors, and this
category is actually a Grothendieck topos. The comparison functor
$I:\Nom\to[\ibb,\Set]$ has a left adjoint $I^*$. We know that $I$
preserves filtered colimits and all limits, while $I^*$ preserves
finite limits and all colimits.

\medskip\noindent\textbf{Abstraction.} Let $(X,\cdot)$ be a
nominal set. We consider the set $[\ncal]X$ consisting of
equivalence classes of pairs $(a,x)\in\ncal\times X$ for the
equivalence relation $\sim$ given by $(a,x)\sim (b,y)$ if and only
if there exists $c\in\ncal\setminus\{a,b\}$, such that $c$ is
fresh for $x$ and for $y$ and $(a\ c)\cdot x=(b\ c)\cdot y$. Let
$[a]x$ denote the equivalence class of $(a,x)$. There is a left
action of $\frak{S}(\ncal)$ on $[\ncal]X$ given by $\pi\circ
[a]x=[\pi(a)]\pi\cdot x$, so the set $[\ncal]X$ can be endowed
with a nominal set structure. In fact, the above construction
extends to a functor $[\ncal]:\Nom\to\Nom$, called
\emph{abstraction} or $\ncal$-abstraction in
\cite{gabb-pitt:lics99}.

\medskip\noindent We have a similar notion of abstraction on $[\ibb,\Set]$,
given by a functor $\delta:[\ibb,\Set]\to[\ibb,\Set]$ defined in
Figure~\ref{fig:delta-UA}. As one might expect, $[\ncal]$ and
$\delta$ are related to each other via the adjunction $I^*\dashv
I$, see Section~\ref{sec:transport}.

\section{Finitary based and uniform
monads}\label{sec:fb-and-uni-mon}

\noindent The aim of this section is two-fold: First, to study monads on $\Nom$. Second, to show how to transport monads from $\Nom$ to $[\ibb,\Set]$. The category theoretic analysis is simplified by abstracting from (\ref{equ:nom-triangle}) and studying instead
\begin{equation}\label{equ:KX}
  \xymatrix{\K^\lbb\ar@/_.7pc/[rr] \ar[dd]  &\bot& (X^\tbb)^\mbb\ar@/_.7pc/[ll] \ar[dd]\\
    \\
    \K
    \ar@/_.7pc/[rr]_{I}\ar@/^.7pc/[ddr]^{U}\POS!R(-.7),\ar@(ul,dl)_{\lbb}
    &\bot
    & \X^\tbb\ar@/_.7pc/[ll]_{I^*}\ar@/^.7pc/[ddl]^{U^\tbb}
      \POS!R(.7),\ar@(ur,dr)^{\mbb}\\
    & & \\
    & \X\ar@/^.7pc/[uur]^{F^\tbb}\ar@/^.7pc/[uul]^{F}
      \POS!R(.7),\ar@(ur,dr)^{\tbb}
    &}
\end{equation}
where $\K$ and $\X^\tbb$ replace $\Nom$ and $[\ibb,\Set]$. $\lbb$ and $\mbb$ are monads, $\K^\lbb$ and  $(X^\tbb)^\mbb$ are the associated categories of algebras.

\medskip\noindent Our assumptions are the following. $\X$ and $\K$ are
locally finitely presentable (l.f.p.) categories \cite{ar} and
$F\dashv U:\K\to\X$ is a finitary adjunction of descent type. This
means that the comparison functor $I:\K\to\X ^\tbb$ is full and
faithful, where $\tbb$ is the monad generated by the adjunction.
Equivalently, $F\dashv U$ is of descent type if every commutative
diagram
\begin{equation}
\label{eq:coeq} \vcenter{ \xymatrix{ FUFUA \ar @<.5ex>
[0,1]^-{\eps FUA} \ar @<-.5ex> [0,1]_-{FU\eps A} & FUA
\ar[0,1]^-{\eps A} & A } }
\end{equation}
is a coequalizer, where $\eps$ denotes the counit of $F\dashv U$.

\medskip\noindent The main contribution of this section is a notion of
functors/monads that can comfortably be transported back and forth
from $\K$ to $\X^\tbb$ using the above adjunction $I^*\dashv I$.
These are exactly those functors/monads that are determined by
their behaviour on finitely generated free objects. They can be
presented by finitary signatures of a special kind: the only
admissible arities are objects, free on finitely presentable
(f.p.) objects of $\X$. As we will see in the next section, this
means that they can be presented by operations and equations in
the sense of universal algebra.

\medskip\noindent We recall first what is meant by signatures and equational
presentations in category theory.

\subsection{Finitary (based) signatures}

\noindent
In \cite{kell-powe:adjunctions}, Kelly and Power proved that finitary
monads on a general l.f.p.\ category $\K$ indeed capture the idea of
equational presentations of algebras on $\K$.  Moreover, the monadic
approach coincides with the UA-approach described in Section~\ref{sec:preliminaries}
in case when $\K=\Set^\Srt$ where $\Srt$ is a set (of sorts). That is,
the presentation (in the sense of Kelly and Power) of any finitary
monad $\tbb$ on $\Set^\Srt$ is a UA-presentation, i.e., $(\Set^\Srt)^\tbb$
is equivalent to a many-sorted variety in the sense of universal
algebra. Figure~\ref{fig:SetI} shows such a presentation where $\tbb$
is as in (\ref{equ:nom-triangle}).

\medskip\noindent The concept of an equational presentation in a general
l.f.p.\ category generalizes the triad
\begin{center}
{\small finitary signatures, terms of depth $\leq 1$, equational theories}
\end{center}
of universal algebra on (many-sorted) sets to the triad
\begin{center}
{\small finitary signatures, finitary endofunctors, finitary monads}
\end{center}
of category theory.

\medskip\noindent The important ingredient of the presentation result of
Kelly and Power \cite{kell-powe:adjunctions} is the recognition of
properties of the adjunction between the elements of the above
triad: for every finitary monad $\tbb$ on $\K$, there exist two
finitary signatures $\Gamma$ and $\Sigma$ and a coequalizer
diagram
$$
\xymatrix{\fbb_\Gamma \ar @<.5ex> [0,1] \ar @<-.5ex> [0,1] &
\fbb_\Sigma \ar[0,1] & \tbb }
$$
in the category of finitary monads on $\K$, where $\fbb_\Gamma$
and $\fbb_\Sigma$ are free (finitary) monads on $\Gamma$ and
$\Sigma$, respectively. A \emph{finitary signature\/} $\Sigma$ on
$\K$ is a family $\Sigma n$ of objects of $\K$ indexed by f.p.
objects $n$ in $\K$. Similarly for $\Gamma$.

\medskip\noindent In fact, the above coequalizer expresses exactly the fact
that $\tbb$-algebras are precisely those $\Sigma$-algebras
satisfying equations specified by the parallel pair. We refer the
reader to \cite{kell-powe:adjunctions} for more details.

\medskip\noindent In what follows, a {\em special kind\/} of finitary
signature on $\K$ will prove to be useful:

\begin{definition}
  Given an adjunction $F\dashv U:\K\to\X$ of descent type,
  an \emph{fb-signature\/} on $\K$ is a
  family $\Sigma n$ of objects of $\K$, indexed by f.p.\  objects
  $n$ in $\X$.
\end{definition}

\noindent Notice that every object of the form $Fn$ is f.p. in
$\K$. Hence fb-signatures are exactly those finitary signatures on
$\K$ that have ``nonempty'' objects of operations only for arities
of the form $Fn$, $n$ f.p. in $\X$. That is, as opposed to
finitary signatures, fb-signatures take arities in $\X$ instead of
$\K$.

\subsection{Finitary and based functors/monads}

\begin{figure}
\fbox{\parbox{\columnwidth}{ arities in $[|\ibb|,\Set]_\fp$:
\par\quad $N_{S,a}=|\ibb|(S\cup\{a\},-)$ for $S\subseteq_{\mathit{f}}\ncal$ and $a\not\in S$.

\medskip fb-signature: \par\quad $\Sigma_\delta:
[|\ibb|,\Set]_\fp\to [\ibb,\Set]$
\par\quad $\Sigma_\delta(N_{S,a})=\ibb(S,-)$, empty
otherwise.

\medskip polynomial functor induced by the signature:
\par\quad $H_{\Sigma_\delta}:[\ibb,\Set]\to[\ibb,\Set]$ given as
$H_{\Sigma_\delta}=\Lan{F}{\Sigma_\delta}$
\par\quad $H_{\Sigma_\delta}(X) = \coprod\limits_{N_{S,a}}
X(S\cup\{a\})\bullet \ibb(S,-)$.

\medskip equations omitted (but see Figure~\ref{fig:delta-UA})
}} \caption{Kelly-Power (KP) presentation of $\delta$
} \label{fig:delta}
\end{figure}

\noindent
The functorial counterpart of fb-signatures is the following
notion:

\begin{definition} A functor $L:\K\to\K$ is called {\em based\/} if
  $L$ preserves all coequalizers of type (\ref{eq:coeq}). A monad
  $\mbb=(M,\mu,\eta)$ on $\K$ is called based if $M$ is a based
  functor. A finitary and based functor/monad is called an
  \emph{fb-functor/monad}.
\end{definition}

\begin{remark}
It can be proved that fb-endofunctors of $\K$
are exactly those that are determined by their
values on objects of the form $Fn$,
where $n$ is f.p.\ in $\X$.
\end{remark}

\noindent Let $\fbEnd{\K}$ denote the full subcategory of
$[\K,\K]$ consisting of fb-functors, and let $\fbMnd{\K}$ denote
the category of fb-monads on $\K$. Any fb-monad on $\K$ can be
presented by operations taking arities from finitely presentable
objects of $\X$. To make this precise:

\begin{theorem} An fb-functor/monad on $\K$ can be
  presented by operations taking arities from f.p. objects of $\X$. Conversely, if a
  monad has such a presentation then it is finitary based.
\end{theorem}

\begin{remark}
  Since any fb-functor/monad is {\em a fortiori\/} finitary, it can be
  equationally presented in the sense of Kelly and
  Power~\cite{kell-powe:adjunctions} using arities from $\K_\fp$.
  The import of the above result is that arities are
  ``finitely generated'' free objects $Fn$. Therefore, one can work with arities $n$ which are f.p. in $\X$.
\end{remark}

\noindent We can
apply all the above results to endofunctors/monads on $\X^\tbb$.
Fb-endofunctors on $\X^\tbb$ are exactly those that are {\em
  determined by values on finitely generated free algebras\/}, since
based now means relative to the monadic adjunction $F^\tbb\dashv
U^\tbb:\X^\tbb\to\X$.

\begin{example}
  The presentation of the abstraction functor from
  Section~\ref{sec:preliminaries} is given in Figure~\ref{fig:delta}
  and, using the notation from universal algebra, in
  Figure~\ref{fig:delta-UA}.
\end{example}

\noindent The following two results will be used in the
Section~\ref{sec:ua-SetI} to show that fb-monads have
presentations in the sense of universal algebra. In a slogan,
these results show that fb-monads are `universal algebraic'.
\begin{proposition}\label{prop:fb=sifted}
  Suppose $U:\X^\tbb\to\Set^\Srt$ is a many-sorted variety.  An
  endofunctor/monad on $\X^\tbb$ is finitary based iff it preserves
  sifted colimits\footnote{For an introduction to sifted colimits see
  \cite{ar:sifted-colim}.}.
\end{proposition}

\begin{theorem}[monadic composition theorem]\label{thm:monadic-composition}
Suppose that $\tbb$ is a finitary monad on an l.f.p.\ category
$\X$ and $\mbb$ an fb-monad on $\X^\tbb$. Then the composite
$$
\xymatrix{(\X^\tbb)^\mbb \ar[0,1] & \X^\tbb \ar[0,1] & \X }
$$
of the forgetful functors is monadic.
\end{theorem}

\subsection{Transporting monads and algebras}\label{sec:transport}

\noindent Since fb-functors are exactly those determined by values on
``finitely generated'' free objects, they have nice properties
w.r.t. transport back and forth along the adjunction $I^*\dashv
I$. The reason for their nice behaviour is, essentially, that $I$ is a
comparison functor and such functors interact nicely with free
objects.

\begin{theorem}
  The assignment $L\mapsto IL I^*$ constitutes a functor
  $\Phi:\fbEnd{\K}\to\fbEnd{\X^\tbb}$ that lifts to a functor
  $\widehat{\Phi}:\fbMnd{\K}\to\fbMnd{\X^\tbb}$.  Both $\Phi$ and
  $\widehat{\Phi}$ are full, faithful and have left adjoints.
  The left adjoint of $\Phi$ is given by $W\mapsto I^*WI$.
\end{theorem}

\begin{example}\label{exle:transport}
  $\delta$ and $[\ncal]$, as well as polynomial functors are
  transported to each other.
\end{example}

\noindent Next, we consider the effect of transport on algebras.
It turns our that the adjunction $I^*\dashv I$ lifts to an
adjunction between the categories of algebras.

\begin{theorem}\label{thm:transf-alg}
  Consider a fb-functor/monad $L$ on $\K$ and let $M=ILI^*$ be its
  ``transport along $I$''. Then there are diagrams
\begin{equation*}
  \xymatrix{
    \textit{L-Alg}\ar[r]^{K} \ar[d]
    &\textit{M-Alg} \ar[d]\\
    \K \ar[r]_{I} \POS!R(-.7),\ar@(ul,dl)^{L}
    & \X^\tbb \POS!R(.7),\ar@(ur,dr)_{M}\\
  }\ \ \ \ \
  \xymatrix{
    \textit{L-Alg}\ar[d]
    &\textit{M-Alg}\ar[l]_{K^*} \ar[d]\\
    \K \POS!R(-.7),\ar@(ul,dl)^{L}
    & \X^\tbb\ar[l]_{I^*} \POS!R(.7),\ar@(ur,dr)_{M}\\
  }
\end{equation*}
commuting up to isomorphism, the left-hand one being a
pseudopullback. Moreover, $K^*\dashv K$ holds.
\end{theorem}

\medskip\noindent Pseudopullbacks are a ``bicategorical'' notion of
pullbacks. The pseudopullback condition means that every
$M$-algebra with carrier from $\K$ is an $L$-algebra. This will be
used in Section~\ref{sec:ua-nom}.

\subsection{Uniform monads}\label{sec:unif-monads}

\noindent An important feature of nominal sets, but also other categories for variable binding \cite{tana-powe:variable-binding} is the presence of an abstraction functor, say $D$. It is therefore of interest to study functors (monads) $H$ which have the property that $D$ lifts to $H$-algebras, that is, there is a `distributive law' $H D\to D H$: Given $HA\to A$ we obtain an $H$-algebra $HD A\to DHA\to DA$ over $DA$.

\medskip\noindent From now on, we instantiate $\K$ in (2) with $\Nom$, hence
$D$ is either $[\ncal]$ or $\delta$ as in
Section~\ref{sec:preliminaries}. We leave a more general
development for future work.

\begin{definition}\label{def:uni-funct}
An endofunctor $H$ on $\Nom$ (or $[\ibb,\Set]$) is called
\emph{uniform} if there exists a natural transformation
$H[\ncal]\to[\ncal] H$ (or $H\delta\to\delta H$).
\end{definition}

\begin{example}
Polynomial functors and $\delta$ are uniform. Figure~\ref{fig:non-uniform-ua} shows an fb-functor that is not uniform.
\end{example}

\medskip\noindent In the case of monads, the natural transformation
needs to satisfy an additional property and is then called a
distributive law \cite{johnstone:adjoint-lifting}.

\begin{definition}\label{def:uni-cat-mon}
  A monad on $\Nom$, respectively on $[\ibb,\Set]$, is called uniform if
  it has a distributive law over $[\ncal]$, respectively over
  $\delta$.
\end{definition}

\begin{example}
 $\delta$ is uniform. In
Figure~\ref{fig:non-uniform-ua} we describe a fb-functor that is not
uniform.
\end{example}

\noindent This allows us to define abstraction of algebras. We
spell it out for $\delta$ and uniform functors, the remaining
cases are analogous.

\begin{definition}\label{def:abstraction}
Suppose $H$ is a uniform functor by means of a distributive
law $\tau:H\delta\to\delta H$. Then the abstraction of
an $H$-algebra $(A,a)$ is an $H$-algebra $(\delta A,Ha\circ\tau_A)$.
\end{definition}

\begin{proposition}
If an fb-functor/monad $L$ on $\Nom$ distributes over $[\ncal]$,
then the transport
$M$ along $I$ distributes over $\delta$. Conversely, if $M$ distributes over $\delta$ then $[\ncal]$ distributes over $L$.
\end{proposition}

\section{Universal algebra over $[\ibb,\Set]$}\label{sec:ua-SetI}
%
%
\begin{figure}
  \fbox{\parbox{\columnwidth}{ operation symbols $\Op_{[\ibb,\Set]}$:
 \[\begin{array}{ll}
    (b/a)_S:S\cup\{a\}\to S\cup\{b\} & a\not = b,\ a\not\in S,\
    b\not\in
    S \\
    w_{S,a}:S\to S\cup\{a\} & a\not\in S
  \end{array}
  \]
  equations $E_{[\ibb,\Set]}$:
  \[
  \begin{array}{lll} (a/b)_S(b/a)_S(x)= x\ \\
    (b/a)_{S\cup\{d\}}(d/c)_{S\cup\{a\}}(x)=(d/c)_{S\cup\{b\}}(b/a)_{S\cup\{c\}}(x)\ \\
    (c/b)_S(b/a)_S(x)=(c/a)_S\ \\
    (b/a)_{S\cup\{c\}}w_{S\cup\{a\},c}(x)=w_{S\cup\{b\},c}(b/a)_S\ \\
    (b/a)_Sw_{S,a}(x)=w_{S,b}(x)\ \\
    w_{S\cup\{b\},a}w_{S,b}(x)=w_{S\cup\{a\},b}w_{S,a}(x)\
  \end{array}
  \]
 }}
\caption{UA-theory of  $[\ibb,\Set]$
} \label{fig:SetI}
\end{figure}

\noindent In this section we see that fb-monads on $[\ibb,\Set]$
are given by universal algebra (UA) theories on $[\ibb,\Set]$.
Corresponding to the concept of uniform monad we introduce the
notions of uniform signature, uniform equations and uniform
UA-theories. Similar to Birkhoff's variety theorem, we can
characterise classes of algebras definable by uniform equations as
those that are closed under images, subalgebras, products and
abstraction. We also prove the uniform analogue of the
quasivariety theorem.

\subsection{Equational theories}
\noindent As explained in Section~\ref{sec:preliminaries}, our
notions of many-sorted signature $(\Srt,\Op)$, equational theory
$\langle\Srt,\Op,E\rangle$, algebras $\Alg{\Srt,\Op,E}$ are those
of Universal Algebra.  We are interested in $\Srt=|\ibb|$.
Referring to Figure~\ref{fig:SetI}, we call

\begin{equation}\label{eq:ThOverI}
  \langle |\ibb|\ ,\ \Op_{[\ibb,\Set]}\uplus
  \Op\ ,\ E_{[\ibb,\Set]}\uplus E\rangle
\end{equation}

\noindent a \emph{theory over $\ibb$}. If equations in $E$ do not
contain nested occurrences of operation in $\Op$ we say that the
theory is of \emph{rank 1}, see Figure~\ref{fig:delta} for an
example.

\begin{proposition}[\cite{kurz-rosi:strcompl,kurz-petr:cmcs08-j}]
\label{prop:pres-fun} A theory $\langle\Srt,\Op,E\rangle$ over
$\ibb$ of rank 1 determines a functor
$M:[\ibb,\Set]\to[\ibb,\Set]$. Moreover,
$\Alg{M}\cong\Alg{\Srt,\Op,E}$.
\end{proposition}

\noindent In one-sorted universal algebra such a functor is
typically a polynomial functor
$X\mapsto LX=\coprod_{n\in\nbb} \Set(n,X)\bullet\Sigma n$, where
$\Set(n,X)\bullet\Sigma n$ denotes the coproduct
of $\Set(n,X)$-many copies of $\Sigma n$.
Hence $\Sigma n$ is the set of $n$-ary operations.
Here, apart from polynomial
functors, we are also interested in functors specifying operations
involving binders, the most basic one being the $\delta$ of
Figure~\ref{fig:delta-UA}.

\noindent Specifying additional operations by a functor has the
advantage that the initial algebra of terms comes equipped with an
inductive principle.  For an example see how $\lambda$-terms form
the initial algebra for a functor in
\cite{fpt:lics99,gabb-pitt:lics99,hofmann:lics99}.

\subsection{Relating KP- and UA-presentations }
 \noindent We argue that the fb-monads from
Section~\ref{sec:fb-and-uni-mon} are precisely those monads that
have a UA-presentation.

\begin{figure}
\fbox{\parbox{\columnwidth}{ operation symbols $\Op_\delta$:
\par \
    \par\quad $[a]_S : S\cup\{a\}\to S$
    \par\quad for all finite sets $S$ and $a\notin S$

\medskip\noindent equations $E_\delta$:
\[\begin{array}{ll}
    (c/b)_S[a]_{S\cup\{b\}}t = [a]_{S\cup\{c\}}
    (c/b)_{S\cup\{a\}}t &\   t:S\cup\{a, b\}  \\
    {[a]_{S}t}=[b]_{S}(b/a)_St &\ t:S\cup\{a\} \\
    w_{S,b}[a]_{S}t=[a]_{S\cup\{b\}}w_{S\cup\{a\},b} t & \ t:
    S\cup\{a\}
  \end{array}
\]}} \caption{UA-presentation of $\delta$
}\label{fig:delta-UA}
\end{figure}

\begin{example}\label{exle:non-uniform-ua}
  Consider a UA-signature as in (\ref{eq:ThOverI}) with $\Op$ containing one
  operation $\app: \emptyset,\emptyset\to \emptyset$ and
  $E=\emptyset$.
  Consider $N:|\ibb|\to\Set$ defined as $N(\emptyset)=2$ and empty
  otherwise.
  The corresponding fb-signature
  $\Sigma: |[|\ibb|,\Set]_\fp|\to [\ibb,\Set]$ maps all f.p.
  objects in $[|\ibb|,\Set]$
  to 0 with the exception of $N$ which is mapped
to $\ibb(\emptyset,{-})$ . The endofunctor presented by $\Sigma$
then is $H_{\Sigma}(X)= (X(\emptyset)\times X(\emptyset))\bullet
\ibb(\emptyset,-)$. Going back from $H_\Sigma$ to a
UA-presentation gives us the theory of
Figure~\ref{fig:non-uniform-ua}. This theory is different from the
one we started with, but the two theories are equivalent: they
define isomorphic categories of algebras.
\end{example}

\noindent This example can be generalised and similar to
Proposition~\ref{prop:pres-fun} we have

\begin{proposition}\label{prop:pres-mon}
  Every UA-theory over $\ibb$ gives rise to an fb-monad on $[\ibb,\Set]$.
\end{proposition}

\noindent Conversely, fb-functors/monads have UA-presentations.
\begin{theorem}
Every fb-functor on $[\ibb,\Set]$ has a presentation as a
UA-theory over $\ibb$ of rank 1.
\end{theorem}
This is a consequence of Proposition~\ref{prop:fb=sifted} and \cite{kurz-rosi:strcompl,kurz-petr:cmcs08-j}.

\begin{theorem}
Every fb-monad on $[\ibb,\Set]$ has a presentation as a UA-theory
over $\ibb$.
\end{theorem}
\noindent This is a consequence of
Theorem~\ref{thm:monadic-composition}.

\subsection{Uniform UA-theories}\label{sec:unif-ua-th}

\noindent Let us give an intuitive motivation for the notions
introduced in this section. Assume we want to investigate
algebraic theories over nominal sets by studying their transport
to $[\ibb,\Set]$. Suppose we have some notion of signature and
equations over nominal sets, such as the nominal logics of
\cite{gabb-math:nomuae,clou-pitt:nom-equ-log}. A nominal set $X$
satisfies an equation, if for any instantiation of the variables,
possibly respecting some freshness constraints, we get equality in
$X$. Notice that the support of the elements of $X$ used to
instantiate the variables can be arbitrarily large. Let us think
what this means in terms of the corresponding presheaf $IX$. For a
finite set of names $S$, $IX(S)$ is the set of elements of $X$
supported by $S$.
 So $IX$ should satisfy not
one, but a set of `uniform' equations, (for an example, see
Figure~\ref{fig:eta}). This means that we should be able to extend
in a `uniform' way the operation symbols together with their
arities, the sort of the equations and the sort of the variables.
We formalize this below, following the same lines as in
\cite{kurz-petr:domains9}. Moreover, we prove that this concrete
syntax implements the notions introduced in Section
\ref{sec:unif-monads}.

\begin{definition}\label{def:uni-UA-sig} A  UA-signature over $\ibb$ of the form $\langle |\ibb|\ ,\ \Op_{[\ibb,\Set]}\uplus
  \Op\ \rangle$ is called uniform if the set $\Op$ of operation symbols can be organized as a
  presheaf, abusively also denoted by
$\Op\in[\ibb,\Set]$, such that any operation symbol $f\in\Op(S)$
has arity of the form

$$f:S_1,\ldots, S_n\to S_0$$

\noindent with $\cup S_i =S$. Additionally, we require that for
any injective map $u:S\to T$ the operation symbol $\Op(u)(f)$ has
arity
$$\Op(u)(f): T\setminus u[S\setminus S_1],\ldots,
T\setminus u[S\setminus S_n]\to T\setminus u[S\setminus S_0]$$
\noindent where $u[S\setminus S_i]$ denotes the direct image of
$S\setminus S_i$ under $u$.  For simplicity let $u\cdot f$ denote
$\Op(u)(f)$.
\end{definition}

\noindent The intention here is that $S\setminus S_i$ is the set
of names bound by $f$ at the corresponding position. For example,
the operations in Figure~\ref{fig:delta-UA} form a uniform
signature. They can be structured as a presheaf as follows:
    \begin{equation}\label{equ:delta-uni}
    \begin{array}{l}
      {[a]}_S\in\Op(S\cup\{a\})\\
      w_b\cdot [a]_S =[a]_{S\cup\{b\}}\\
      (b/a)_S\cdot [a]_S = [b]_S
    \end{array}
    \end{equation}

\medskip\noindent For such a signature we define the notions of uniform
term and uniform equation. The intuition here is that a uniform
equation generates a set of equations in the sense of universal
algebra.

\medskip\noindent A \emph{uniform term} \ $t:T$
  \ for a uniform signature is a term $t$ of type $T$ formed according
  to the rules in Figure~\ref{fig:uniform-terms}. Each rule can be
instantiated in an infinite number of ways: $T$ ranges over finite
sets of names and $a,b$ over names. The notation $T\uplus\{a\}$
indicates that an instantiation of the schema is only allowed for
those sets $T$ and those atoms $a$ where $a\not\in T$. A
\emph{uniform
    equation} is a pair of uniform terms of the same sort $u = v:T$,
  such that any variable $X$ appears with the same type $T_X$ in both
  $u$ and $v$. A \emph{uniform theory} consists of a set of uniform
  equations.

  \medskip\noindent A uniform equation  $u=v:T$ is not an equation in the
sense of universal algebra, but it generates a set of equations
indexed over all finite sets of names $S$ that are disjoint from
$T$. We will call these equations the translations of $u=v:T$ by
$S$, and they are defined below. These translations should involve
enlarging the sort of the variables. However this is not always
possible, for example if we have a subterm $w_a X$ of an equation,
then the sort of $X$ cannot contain the name $a$.

\begin{definition}\label{def:fresh-uni-equ} The freshness set of a variable $X$
  appearing with sort $T_X$ in an equation $E$ of the form $u=v:T$ is
  the set
$$\Fr_E(X)=\bigcup\limits_{t:T}T\setminus T_X$$ where the union is
taken over all sub-terms $t$ of either $u$ or $v$ that contain the
variable $X$.
\end{definition}

\begin{example}\label{exp:fresh} As an example, let us consider a set of operation
symbols
\[\begin{array}{l}
  \ a_S\ :\ S\cup\{a\}\\
  \ \app_S\ :\ S, S\to S\\
  \ [a]_{S}\ :\ S\cup\{a\}\to S
\end{array}\]
\noindent In fact these operations subject to some equations give
a presentation for  the functor $LX=\ncal+\delta X+ X\times X$,
whose initial algebra is the presheaf of $\alpha$-equivalence
classes of $\lambda$-terms, see \cite[Section
4]{kurz-petr:domains9} for details on this.

\medskip\noindent For the uniform equation
$[a]_{\emptyset}\app_{\{a\}}(w_aX,a_\emptyset)=X$ the freshness
set of $X$ is $\Fr(X)=\{a\}$. In Figure~\ref{fig:eta} we see that
this equation corresponds to equations in other nominal logics
having as side condition that $a$ is fresh for $X$.
\end{example}

\newcommand{\trans[1]}{\mathit{tr}_{\!#1}\,}

\begin{definition}\label{def:trans-equ} The translation of an equation
  $E$ of the form $u=v:T_E$ by a name $a\not\in T_E$
  is  an equation $\trans[a](u)=\trans[a](v)$ of sort $T\cup \{a\}$, where the translation $\trans[a](t:T)$ (with $a\not\in T$) of a sub-term
  $t$ of either $u$ or $v$ is defined recursively by

  \begin{equation}\label{equ:trans-st-univ-alg}
    \begin{array}{l}
    \trans[a](f(t_1,\dots,t_n):T_0)\stackrel{(a\not\in T)}{=}(w_a\cdot f)(\trans[a](t_1),\dots,\trans[a](t_n)) \\
    \trans[a](f(t_1,\dots,t_n):T_0)\stackrel{(a\in T)}{=}w_a(f(t_1,\dots,t_n))  \\
    \\
    \trans[a](w_b t: T\uplus\{b\})=w_{S\cup \{a\},b}\,\trans[a](t:T) \\
\\
    \trans[a]((b/c)t:T\uplus\{b\})\stackrel{(a\neq c)}{=}(b/c)_{T\cup\{a\}}\,\trans[a](t:T\uplus\{c\})\\
    \trans[a]((b/a)t:T\uplus\{b\})\ =\ w_a (b/a)_{T}\,t  \\
 \\
    \trans[a](X:T_X)= w_aX_{
        T_X}  \quad\mathrm{if\ } a\in \Fr_E(X)\\
    \trans[a](X:T_X)=X'_{
        T_X\cup \{a\}}  \quad\mathrm{if\ } a\not\in \Fr_E(X)\\
    \end{array}
  \end{equation}

   \noindent where in the first two conditions $f: T_1,\dots, T_n\to T_0$ is an operation symbol in $\Op$. In the last condition
  $X'_{ T_X\cup
    \{a\}}$ is a variable of sort $ T_X\cup \{a\}$.

    The translation of an equation
  $E$ of the form $u=v:T_E$ by a set $S=\{a_1,\dots,a_k\}$ disjoint from $T_E$ is a
  defined as $\trans[a_1](\dots\trans[a_k](u=v:T)\dots ): T\cup S$. (The
  chosen order of the elements of $S$ is irrelevant).
\end{definition}

\medskip\noindent We will say that a set of (standard universal
algebra) equations is \emph{uniformly generated} by a uniform
theory $\ucal$ if it consists of all possible translations of the
uniform equations in $\ucal$.

\begin{example}
The UA-theory expressing the eta-equivalence of the
$\lambda$-calculus is uniformly generated by the uniform equation
of the last line of Figure~\ref{fig:eta}.
\end{example}

\begin{definition}\label{uni-UA-th} A uniform UA-theory over $\ibb$ is a theory
$ \langle |\ibb|\ ,\ \Op_{[\ibb,\Set]}\uplus
  \Op\ ,\ E_{[\ibb,\Set]}\uplus E \uplus E_\Op\rangle$  such that the set of equations $E$ is uniformly generated by a
uniform theory and $E_\Op$ is the set of equations of the form:
\[
 \begin{array}{l}
 (w_a\cdot f)(w_ax_1,\dots,w_ax_n)=w_af(x_1,\dots,x_n)\\
 ((a/b)_{S\setminus\{b\}}\cdot
      f)(\langle a/b \rangle_{S_1\setminus\{b\}}x_1,\dots,\langle
      a/b\rangle_{S_n\setminus\{b\}}x_n)= \\
      $$\langle a/b\rangle_{S_0\setminus\{b\}}f(x_1,\dots,x_n)
  \end{array}
  \]

    \noindent for  $f\in\Op(S)$ having arity $S_1,\ldots, S_n\to S_0$,
   $a\not\in S$ and $b\in S$, with the additional convention that $\langle a/b\rangle_{S_i\setminus \{b\}}$
   denotes the identity on $S_i$ if $b\not\in S_i$ and
   $(a/b)_{S\setminus\{b\}}$ if $b\in S_i$.
\end{definition}

\noindent Next, we will see that there is a strong connection
between uniform UA-theories and the concept of abstraction. The
reason for this is the existence of an isomorphism for every
finite set $S$ and $a\notin S$
$$A(S\cup\{a\})\cong\delta A(S)$$
\noindent that maps $x\in A(S\cup\{a\})$ to $[a]_Sx$.

\medskip\noindent Consider a uniform signature as in Definition \ref{def:uni-UA-sig} and
let $A$ be an algebra for the uniform theory $ \langle |\ibb|\ ,\
\Op_{[\ibb,\Set]}\uplus
  \Op\ ,\ E_{[\ibb,\Set]}\uplus E_\Op\rangle$. We can define the
  \emph{abstraction} of $A$  to be an algebra with carrier $\delta A$
  and the interpretation of an operation symbol in $\Op(S)$ of the form $f:S_1,\dots,S_n\to S_0$ given by:
  $$f^{\delta A}([a]_{S_1}x_1, \ldots, [a]_{S_n}x_n)=[a]_{S_0}(w_a\cdot f^A)(x_1,
  \ldots,
   x_n)$$
 \noindent for some  $a\notin S$.

 \medskip\noindent The next proposition is based on the observation
 that an algebra $\delta A$ satisfies an equation $E$ if and only
 if the algebra $A$ satisfies the translation $\trans[a]{E}$ of an equation by a
 \emph{new} name $a$.

 \begin{proposition}\label{prop:clos-abs} A class of algebras for a uniform UA-theory $ \langle |\ibb|\ ,\
\Op_{[\ibb,\Set]}\uplus
  \Op\ ,\ E_{[\ibb,\Set]}\uplus E_\Op\rangle$
 defined by uniform equations $E$ is closed under abstraction.
 \end{proposition}

\noindent From this it follows that a class of algebras for a
uniform UA-theory defined by additional uniform equations is
closed under abstraction. This means that the abstraction functor
$\delta$ lifts to a functor $\tilde{\delta}$ on the categories of
algebras for a uniform UA-theory. Therefore, similar to
Proposition \ref{prop:pres-fun} we have:

\begin{proposition} The functor on $[\ibb,\Set]$ determined by a uniform UA-theory of rank
1 is uniform in the sense of Definition \ref{def:uni-funct}.
\end{proposition}

\begin{example} The functor $\delta$ has a uniform presentation given in Figure \ref{fig:delta-UA}.
  As a counterexample, consider the functor $L_0$ presented in
  Figure \ref{fig:non-uniform-ua}. Although the operations can be
  structured as a presheaf, the presentation is not uniform.
\end{example}

\newcommand{\SetI}{[\ibb,\Set]}

\begin{figure}
\fbox{\parbox{\columnwidth}{
      \[
      \begin{array}{c}
        \rules{t_1:T_1,\ \ldots, \ t_n:T_n}
        {f(t_1,\ldots t_n):T_0}\ (f:T_1,\dots, T_n\to T_0 \in\Op(T))
        \\[1.5em]
        \rules{t:T }
        {w_at:T\uplus \{a\}}
        \quad \quad \rules{t:T\uplus\{a\}}
        {(b/a)t:T\uplus\{b\}}
        \quad \quad \rules{ }
        {X:T_X}
      \end{array}
      \]
    }} \caption{Uniform terms} \label{fig:uniform-terms}
\end{figure}

\begin{figure} \fbox{\parbox{\columnwidth}{ operations:
\smallskip
\par\quad $\app_S:
\emptyset, \emptyset\to S$ for all $S\subseteq_f\ncal$

\medskip equations:
\smallskip
\par \quad $w_a\app_S(x,y)=\app_{S\cup\{a\}}(x,y)$
\par\quad $(b/a)_S\app_{S\cup\{a\}}(x,y)=\app_{S\cup\{b\}}(x,y)$
}} \caption{UA-Presentation of a non-uniform functor
$L_0(X)=(X(\emptyset)\times X(\emptyset))\bullet \ibb(\emptyset, -)$}
\label{fig:non-uniform-ua}
\end{figure}

\noindent Similar to Proposition \ref{prop:pres-mon} we obtain

\begin{proposition}
 Every uniform UA-theory over $\ibb$ gives rise to a uniform fb-monad on
  $[\ibb,\Set]$, see Definition~\ref{def:uni-cat-mon}.
\end{proposition}

\subsection{Results from universal algebra}

\noindent In one-sorted universal algebra, Birkhoff's variety
theorem characterizes equationally definable classes of algebras
as those closed under HSP, that is, homomorphic images,
subalgebras and products. The theorem is not true in general for
many-sorted algebras, see~\cite{arv:varieties}: An equationally
definable class of many-sorted algebras is closed under
homomorphic images, subalgebras, products and \emph{directed
colimits}. However, because of the special structure of the
category $\ibb$, as pointed out in \cite{kurz-petr:domains9}, we
have:

\begin{theorem}\label{thm:hsp}
  Consider a UA-theory over $\ibb$ and let $\acal$ denote its
  algebras. Then a class $\ccal\subseteq\acal$ is equationally
  definable if and only if it is closed under HSP.
\end{theorem}

\noindent There exists a similar characterization of finitary
quasivarieties for many-sorted algebras. These are classes of
algebras definable by implications, where by implication we mean
here a formula
\[ (u_1=v_1)\wedge\dots\wedge(u_n=v_n)\Rightarrow (u_0=v_0) \]
where $u_i=v_i$ are equations. The next theorem is an instance of
the well known quasivariety theorem.

\begin{theorem} Let $\acal$ be the category of algebras for a UA-theory over
$\ibb$. Then a class $\ccal\subseteq\acal$ is implicationally
definable if and only it is closed under subalgebras, products and
filtered colimits.
\end{theorem}

\noindent For the uniform UA-theories we can provide similar
characterizations. The next theorem generalises \cite[Theorem
5.23]{kurz-petr:domains9}. On a category of algebras $\acal$ given
by a uniform UA-theory we have an abstraction operator given by
Proposition~\ref{prop:clos-abs}. We have

\begin{theorem}\label{thm:hspa}
  Consider a uniform UA-theory over $\ibb$ and let $\acal$ denote its
  algebras. Then a class $\ccal\subseteq\acal$ is equationally
  definable by additional uniform equations if and only if it is closed under HSPA, that is, homomorphic images, subalgebras, products and abstraction.
\end{theorem}

\begin{definition} A uniform implication of type $T$ is a formula
\[ (u_1=v_1)\wedge\dots\wedge(u_n=v_n)\Rightarrow (u_0=v_0) :T
\]
\noindent where $u_i=v_i:T_i$ are uniform equations for
$i=0,\dots,n$ and $T=T_0\cup\dots\cup T_n$.
\end{definition}

\noindent Each uniform implication of type $T$ generates a set of
standard universal algebra implications, indexed by finite sets
$S$ with $S\cap T=\emptyset$. We do this by translating each
uniform equation $(u_i=v_i):T_i$ as in
(\ref{equ:trans-st-univ-alg}), with the only difference being that
in the last two relations, we use
$\Fr_{u_0=v_0}(X)\cup\dots\cup\Fr_{u_n=v_n}(X)$ instead of
$\Fr_{u_i=v_i}(X)$.

\medskip\noindent Consider the category of algebras $\acal$ for a uniform
UA-theory. We say that $\ccal\subseteq\acal$ is implicationally
definable by uniform implications if there exits a set of uniform
implications $\ical$ such that $\ccal$ is definable by the set of
UA-implications generated by all the elements of $\ical$. Then we
can prove:

\begin{theorem}\label{thm:quasi}
  Consider a UA-theory over $\ibb$ and let $\acal$ denote its
  algebras. Then a class $\ccal\subseteq\acal$ is implicationally
  definable by uniform implications if and only if it is closed
  under subalgebras, products, filtered colimits and abstraction.
\end{theorem}

\section{Universal algebra over nominal sets}\label{sec:ua-nom}

\noindent Building on the general theory developed in
Section~\ref{sec:fb-and-uni-mon}, we can now transfer properties
and results obtained in universal algebra on $[\ibb,\Set]$ to
nominal sets. To achieve this we use the next theorem, which can
be derived from Theorem~\ref{thm:transf-alg}.

\begin{theorem} Any fb-monad/functor $L$ on $\Nom$ induces a UA-theory $\Phi$ on
$[\ibb,\Set]$, so that the category of $L$-algebras is the
category of $\Phi$-algebras `restricted along $I$'.
\end{theorem}

\medskip\noindent There are several approaches in the literature to
develop algebraic theories over nominal sets: nominal (universal)
algebra of~\cite{gabb-math:nomuae} and NEL
of~\cite{clou-pitt:nom-equ-log}. These approaches fit in the
general framework  developed here, and more importantly, we can
prove new results for them using our technique.

\medskip\noindent For example, the signatures defined
in~\cite{gabb-math:nomuae} are given by functors of the form
$\ncal+[\ncal]+\Sigma$, where $\ncal$ is the constant functor,
$[\ncal]$ is the abstraction functor and $\Sigma$ is a polynomial
functor. These functors are uniform and finitary based. In fact,
in~\cite[Section 6]{kurz-petr:domains9} we have given syntactical
translations of theories of nominal algebra and NEL into uniform
theories, for an example see Figure~\ref{fig:eta}. As anticipated
in Example~\ref{exp:fresh} we translate a freshness condition
$a\#X$  by adding operations symbols of the form $w_a$ in front of
the variable $X$.

\begin{figure}
  \fbox{\parbox{\columnwidth}{
  nominal algebra (\cite{gabb-math:nomuae}):
  \par\quad  $a\#X\vdash [a]\app(X,a)=X$

  \medskip\noindent NEL (\cite[Fig. 4]{clou-pitt:nom-equ-log}):
  \par\quad $a\sfresh x \vdash L_a(A\  x\  V_a)\approx x$

  \medskip\noindent UA-theory:
  \par\quad $[a]_S\app_{S\cup\{a\}}(w_{S,a}X_S,a_S)=X_S$
  \par\quad for all finite $S$, $a\not\in S$ and $X_S$ variable of
  sort S

  \medskip\noindent uniform UA-theory:
  \par\quad $[a]\app_{\{a\}}(w_aX,a)=X$
 }}
\caption{$\eta$-rule for untyped  $\lambda$-calculus}
\label{fig:eta}
\end{figure}

\medskip\noindent In our general setting, we can characterise the
equationally definable subcategories of algebras on nominal sets.
First, let us see what we mean by this.

\begin{definition} Let $L$ be a functor on $\Nom$. A full subcategory $\ccal$ of $L$-algebras is
  equationally definable by a UA-theory $\Phi$ on $\ibb$ if $\ccal$ consists of $L$-algebras $(A,a)$ with
$K(A,a)\models\Phi$, where $K:\textit{L-Alg}\to\Phi\textit{-Alg}$
is the lifting of $I$ as in Theorem~\ref{thm:transf-alg}.
\end{definition}

\noindent The next theorem follows from Theorem~\ref{thm:hsp} and
the observation that a $\Phi$-algebra which lies in  the closure
  under HSP of $I\ccal$ and has as carrier a nominal set is in fact an object of
  $I\ccal$. Here, $I\ccal$ is the subcategory of
  $\Phi\textit{-Alg}$, obtained as the image of $\ccal$ under $I$.

\begin{theorem}
  Let $L$ be a fb-functor/monad on $\Nom$. A class of $L$-algebras is equationally definable if and only if it is
  closed under homomorphic images of support-preserving maps, under
  subalgebras and under products.
\end{theorem}

\begin{remark} We obtain closure under homomorphic images of support-preserving
maps rather than all homomorphic images because $I$ only preserves
the former.
\end{remark}

\noindent But we can do better than that for algebras for a
functor $L$, whose transport on $[\ibb,\Set]$ is given by a
uniform UA-theory of rank 1. In the remainder of this section by
algebras over nominal sets we understand algebras for such
functors. From Theorem~\ref{thm:hspa} we derive

\begin{theorem}
  A class of algebras over nominal sets is definable by uniform equations if and only if it
  is closed under homomorphic images, subalgebras, products, and
  abstraction.
\end{theorem}

\noindent Similarly, using Theorem~\ref{thm:quasi} we can prove a
quasivariety theorem for algebras over nominal sets:

\begin{theorem}
  A class of algebras over nominal sets is definable by uniform implications if and only if it
  is closed under subalgebras, products, filtered colimits and
  abstraction.
\end{theorem}

\noindent These theorems can be transferred to nominal
algebras~\cite{gabb-math:nomuae} and
NEL~\cite{clou-pitt:nom-equ-log}, using the translations given
in~\cite{kurz-petr:domains9}, for example we recover Gabbay's
HSPA-theorem of~\cite{gabbay:hsp}. Additionally we obtain new
results such as:

\begin{theorem} Categories of nominal algebras in the sense
of~\cite{gabb-math:nomuae} are given by uniform monads on $\Nom$.
\end{theorem}

\noindent This is obtained using the fact that the translation of
nominal algebra into uniform theories are semantically invariant
(\cite[Theorem 6.8]{kurz-petr:domains9}), and that uniform
theories are given by uniform fb-monads.

\section{Conclusions }

\noindent We have shown how algebra with variable binding can be
done inside standard many-sorted universal algebra. Our framework
comprises nominal sets as well as the associated presheaf model of
variable binding. Of particular importance here are the results of
Section \ref{sec:transport} which show that universal algebra can
not detect the difference between the two. It also sheds new light
on the different proposals of equational logic for nominal sets
\cite{clou-pitt:nom-equ-log,gabb-math:nomuae}, as they can be
compared now as describing slightly different fragments of the
uniform theories described in Section \ref{sec:unif-ua-th}.

\smallskip Future work:

\smallskip\noindent\textbf{$\bullet$ }
To extend by `uniform implications' the logics
of~\cite{gabb-math:nomuae} and~\cite{clou-pitt:nom-equ-log}.

\smallskip\noindent\textbf{$\bullet$ }  To transfer more results of
universal algebra and to develop applications to the theory of
algebraic data types.

\smallskip\noindent\textbf{$\bullet$ }  To `nominalise' other areas of theoretical
computer science based on universal algebra.

\smallskip\noindent\textbf{$\bullet$ } In particular, there is ongoing work on nominal regular languages and their automata. Appropriate notions of finite algebras are obtained via the named sets of \cite{fmp:fossacs02}.

\smallskip\noindent\textbf{$\bullet$ } Applications to process algebras with name binders. For example, the logic developed in \cite{bons-kurz:lics07} falls into our framework, as do Stark's algebraic models of the $\pi$-calculus \cite{stark:free-alg-pi}.

\smallskip\noindent\textbf{$\bullet$ } Our general aims are related to those of Fiore and Hur~\cite{fior-hur:icalp07}, but instead of developing an abstract framework we focus on particular models and stay inside classical universal algebra. A precise relationship needs to be worked out.

\smallskip\noindent\textbf{$\bullet$ } To extend our framework to other presheaf models of variable binding according to the general theory developed in \cite{tana-powe:variable-binding}.

\smallskip\noindent\textbf{$\bullet$ } To deal with recursion, presheaf models over cpos have been studied in \cite{fms:lics96,stark:lics96}. Let us note that Section~3 as well as \cite{powe-tana:enriched-binding} work in the enriched setting, suggesting to replace $\Set$ by cpos. This raises the interesting question of what `enriched equational logic' is.

\bibliographystyle{plain}

\end{document}